\documentclass[preprint,showpacs,preprintnumbers,amsmath,amssymb,endfloats]{revtex4-2}
\usepackage{graphicx}
\usepackage{dcolumn}
\usepackage{bm}
\usepackage{amssymb}
\usepackage{amsmath}
\usepackage{latexsym}
\usepackage{longtable}
\usepackage{color}

\begin{document}

\title{Self-oscillating  open quantum systems}

\author{V.V.~Sargsyan$^{1,2,a}$, A.A.~Hovhannisyan$^{1,2,b}$, G.G.~Adamian$^{1,c}$,  N.V.~Antonenko$^{1,d}$
}
\affiliation{
$^{1}$Joint Institute for Nuclear Research, 141980 Dubna, Russia\\
$^{2}$A. Alikhanyan National Science Laboratory (YerPhI),  0036 Yerevan, Armenia\\
%
$^{a}$sargsyan@theor.jinr.ru\\
$^{b}$arshakhov@gmail.com\\
$^{c}$adamian@theor.jinr.ru\\
$^{d}$antonenk@theor.jinr.ru
}

\date{\today}

\begin{abstract}
Evolution of occupation number is studied for a bosonic oscillator (with one and two degrees of freedom)
linearly fully coupled  to  fermionic and bosonic heat baths.
The lack of equilibrium in this oscillator is discussed in the light of the creation of an
energy source.
The connection of such a system with known nonlinear self-oscillating systems is shown.

\end{abstract}
\pacs{05.30.-d, 05.40.-a, 03.65.-w, 24.60.-k \\
Key words:   self-oscillating systems, master-equation, non-stationary systems,
 time-dependent  bosonic occupation numbers, energy source}
\maketitle

\section{Introduction}

Many physical problems can be solved by dividing the entire system into an open system, the dynamics of which we are interested in, and one or several thermal baths associated with it
and containing other degrees of freedom \cite{M1,M8,Klim,Zub,M10,Petruccione,Dodonov,Armen,Stefanescu,PA,PRE2020n}. In the case of one bath or several baths of the same statistics and their linear coupling to
the open system, the problems can be solved exactly \cite{PRE1997,PRD1992,PRE2013,PRL2022,JCP2013,PRP1998,PRA2020}.
The problem is getting complicated if the open quantum system is fully coupled to the baths of different statistics \cite{PRE2020n,PhysicaA2019,PRE2020}.
For example, collective motion surrounded by several thermal baths with different statistics occurs when nuclei fuse in a stellar medium.
Indeed, the dynamics in the relevant collective coordinate is related to fusion process.
One thermostat is the nucleon degrees of freedom of the nuclei.
Other thermostats can be associated with  electromagnetic field.
This description allows us to  take effectively  into account the internal structure of interacting nuclei and the properties of   medium.
The formation of atomic (nuclear) molecule   in the external laser  field is another example.
In this case, the electron   (nucleon) degrees of freedom
and the action of the electromagnetic field  can be effectively modeled with
appropriate fermionic and bosonic heat baths, respectively.

In Refs.~\cite{PRE2020n,PhysicaA2019,PRE2020},
it was illustrated that, under certain conditions, a fermionic oscillator (the two-level system) or bosonic oscillator
linearly fully coupled  to several baths of different statistics (fermionic and bosonic)
might never reach the stationary asymptotic limit and that the occupation number  in this system oscillates at large times.
The period of asymptotic oscillations depends on the frequency of the oscillator but not on the complex environment.
This system with a non-stationary asymptotic occupation number can be used as a dynamic or non-stationary  memory storage \cite{PRE2020n} because
the information about the system (population of excitation states and frequency) is preserved at large times
and remains under   external and environmental conditions \cite{PRE2020n}.
For quantum computers, the imposition of unambiguous population of
excited states of the quantum register is also important.
These states relax  to the ground state due to dissipative processes.
Using the systems with   non-stationary asymptotic occupation numbers instead of stationary ones,
one can ensure a sufficient degree of metastability of the excited states of the quantum register  \cite{PRE2020n}.
Note that the non-stationary asymptotic behavior is also known in the self-oscillating chemical reactions \cite{Zaikin}.
The reason for these oscillations is the coupling with various ingradients/steps of the process.

In the present paper, we study the time evolution of occupation numbers of two linearly
coupled bosonic oscillators embedded in the  fermionic and bosonic heat baths
and reveal
the relationship between the non-stationary systems considered and self-oscillating systems.
Note that the systems under consideration  are linearly fully coupled (FC) to the heat baths of different statistics
\cite{M1,M8,Zub,M10,Klim,Petruccione,Armen,Stefanescu,PA,PRE2020n}.

\section{Bosonic oscillator linearly fully coupled with  fermionic and bosonic baths}
The Hamiltonian of total system  (the quantum system plus  two heat baths "$\lambda$", $\lambda=1,2$)  is
written as \cite{PRE2020}
\begin{equation}
\label{ham}
H=H_c+\sum_{\lambda=1,2} H_{\lambda} +  \sum_{\lambda=1,2}H_{c,\lambda},
\end{equation}
where
\begin{equation}
H_c=\hbar \omega a^{\dagger} a
\end{equation}
is the Hamiltonian of the quantum system being bosonic oscillator with frequency $\omega$,
\begin{equation}
\label{eq:baths}
H_{ \lambda}= \sum_i \hbar \omega_{\lambda,i} c^{\dagger}_{\lambda,i} c_{\lambda,i}
\end{equation}
are the Hamiltonians of  heat baths, and
\begin{equation}
\label{eq:int1}
H_{c, \lambda}= \sum_i \alpha_{\lambda,i} (a^{\dagger}+a )(c^{\dagger}_{\lambda,i}+c_{\lambda,i})
\end{equation}
are the interaction Hamiltonians with the  real constants $\alpha_{\lambda,i}$ determining the coupling strengths between bosonic oscillator and
fermionic and bosonic (fermionic-bosonic) heat baths.
For the FC interaction between the  oscillator and heat baths, the  interaction Hamiltonian (\ref{eq:int1}) is  linear in the oscillator and baths operators.
This  has important consequences on the dynamics of the  oscillator by
altering the effective collective potential and by allowing energy to be exchanged with the thermal reservoirs.

The creation  and annihilation   operators of  the oscillator $(a^{\dagger}, a)$ and heat baths $(c^{\dagger}_{\lambda,i}, c_{\lambda,i})$
satisfy the  commutation or anti-commutation relations:
\begin{equation}
\label{eq:perm}
\begin{split}
&aa^{\dagger}- a^{\dagger}a=1,\\
&c_{\lambda,i}c^{\dagger}_{\lambda,i}-\varepsilon_\lambda  c^{\dagger}_{\lambda,i}c_{\lambda,i}=1,
\quad c^{\dagger}_{\lambda,i}c^{\dagger}_{\lambda,i}-\varepsilon_\lambda c^{\dagger}_{\lambda,i}c^{\dagger}_{\lambda,i}=c_{\lambda,i}c_{\lambda,i}-\varepsilon_\lambda  c_{\lambda,i}c_{\lambda,i}=0,
\end{split}
\end{equation}
where  $\varepsilon_\lambda $ are equal to 1 and -1
for the  bosonic and fermionic  heat baths, respectively.

Employing the Hamiltonian (\ref{ham}), we deduce the  equation of motion
for the occupation number operator
\begin{eqnarray}
\frac{d}{dt} {a^\dagger(t)a(t)}&=&
\frac{i}{\hbar}\sum_{\lambda,i}\alpha_{\lambda,i}[a(t)-a^{\dag}(t)][c^{\dagger}_{\lambda,i}(t)+c_{\lambda,i}(t)]\nonumber \\
&=&
\frac{i}{\hbar}\sum_{\lambda,i}\alpha_{\lambda,i}[c^{\dagger}_{\lambda,i}(t)a(t)-a^{\dag}(t)c_{\lambda,i}(t)+
a(t)c_{\lambda,i}(t)-a^{\dag}(t)c^{\dagger}_{\lambda,i}(t)]
\label{master-eq}
\end{eqnarray}
of the bosonic oscillator.
Deriving the  equations of motion
for the operators
$c^{\dagger}_{\lambda,i}(t)a(t)$ and
$a(t)c_{\lambda,i}$ and substituting their formal solutions
in   Eq. (\ref{master-eq})
and taking   the initial conditions $\langle c^{\dagger}_{\lambda,i}(0)a(0)\rangle=\langle a(0)c_{\lambda,i}(0)\rangle=\langle a^{\dag}(0)c_{\lambda,i}(0)\rangle=\langle a^{\dag}(0)c^{\dagger}_{\lambda,i}(0)\rangle=0$
(the symbol $\langle  ... \rangle$ denotes the averaging  over the whole system of heat baths and oscillator), and assuming that
$\langle aa\rangle=\langle a^{\dagger}a^{\dagger}\rangle=\langle c^{\dagger}_{\lambda,i} c^{\dagger}_{\lambda',i'}\rangle =\langle c_{\lambda',i'} c_{\lambda,i}\rangle=0$,
$\langle  c^{\dagger}_{\lambda,i} c_{\lambda',i'}\rangle =\langle c^{\dagger}_{\lambda,i}c_{\lambda,i}\rangle=n^{(\rm\lambda)}_i\delta_{\lambda,\lambda'}\delta_{i,i'}$ (the heat baths consist of  independent oscillators; $n^{(\rm\lambda)}_i$ is the occupation number of the heat bath state $i$), 
and $\langle  a^{\dagger}ac^{\dagger}_{\lambda,i} c_{\lambda,i}\rangle= n_{1} n^{(\rm\lambda)}_i$ (the mean-field approximation),
we obtain the  master-equation for the occupation number $n_{1}=\langle a^{\dag}a\rangle$ of the  bosonic oscillator \cite{PRE2020}:
\begin{eqnarray}
\frac{dn_{1}(t)}{dt}=-2\lambda_1(t)n_{1}(t)+2D_1(t),
\label{eq:namaster2}
\end{eqnarray}
where
\begin{equation}
\label{lambda}
\lambda_1(t)=p\lambda_{\scriptstyle  f}(t)+(1-p)\lambda_{\scriptstyle b}(t) - 2D_{{\scriptstyle  f}}^{(1)}(t)
\end{equation}
and
\begin{eqnarray}
\label{Dif}
D_1(t)=D_{{\scriptstyle  f}}^{(1)}(t)+D_{{\scriptstyle  b}}^{(2)}(t)
\end{eqnarray}
are the time-dependent friction and diffusion coefficients, respectively \cite{PRE2020n}.
The value of $p$, the time-dependent friction
$\lambda_{\scriptstyle f}(t)$ [$\lambda_{\scriptstyle b}(t)$], and partial diffusion $D_{{\scriptstyle  f}}^{(1)}(t)$ [$D_{{\scriptstyle  b}}^{(2)}(t)$]
coefficients for the fermionic [bosonic] oscillator coupled with two fermionic [bosonic] heat baths are given in Appendix A.
The master-equation (\ref{eq:namaster2}) can also be rewritten in the following way:
\begin{eqnarray}
\frac{d^2n_{1}(t)}{dt^2}+2\lambda_1(t)\frac{dn_{1}(t)}{dt}+2\frac{d\lambda_1(t)}{dt}n_{1}(t)=2\frac{dD_1(t)}{dt}.
\label{eq:namaster3}
\end{eqnarray}

In the case of the baths of different statistics the mean-field
approximation is required to obtain the master-equation  (\ref{eq:namaster2}) for the collective subsystem.
As known from nuclear physics, this approximation is good in describing
collective excitations coupled to the internal fermionic degrees of freedom \cite{Peters}.
To derive the master-equation  (\ref{eq:namaster2}), we also assume that
the heat baths are large enough to retain their properties, despite the coupling
with the collective oscillators. So, the usual definition of a heat bath,
as an infinite set
of independent oscillators,
is considered~\cite{PRA2017}.

%
%
%
\begin{figure}
\begin{center}
\includegraphics[scale=0.85,angle=0]{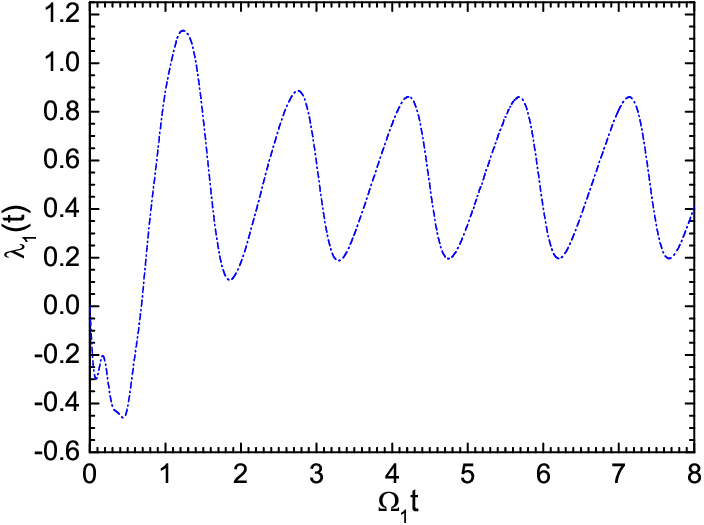}
\includegraphics[scale=0.85,angle=0]{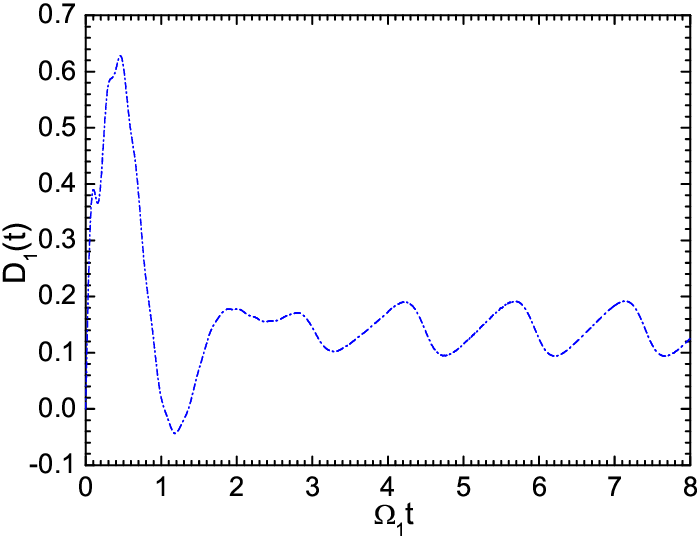}
\includegraphics[scale=1.10,angle=0]{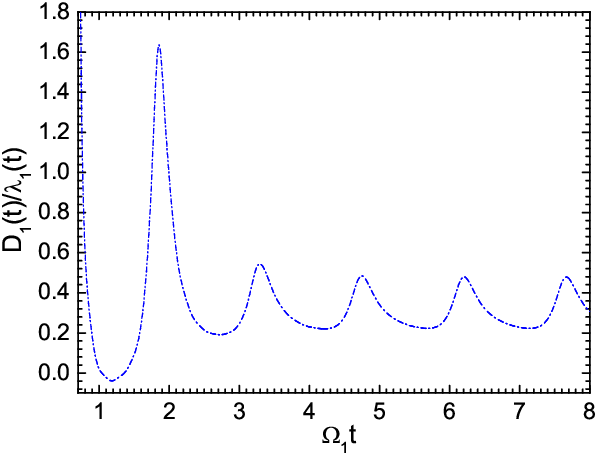}
\end{center}
\vspace{1cm}
\caption{
Calculated dependencies of the friction  $\lambda_1(t)$    and  diffusion
$D_1(t)$  coefficients, and the ratio $D_1(t)/\lambda_1(t)$ on time $t$ for the
 bosonic oscillator.
 The renormalized frequency of the bosonic oscillator is $\Omega_1$ \cite{PRE2020n}.
The fermionic and bosonic baths  have the same
level densities  with the Lorenzian  cut-off parameters (the inverse memory times)  $\gamma_{1}/\Omega_1=10$, $\gamma_{2}/\Omega_1=15$, respectively.
Here the coupling strengths \cite{PRE2020n} between the oscillator and baths are $\alpha_{1}=0.1$ and $\alpha_{2}=0.05$,
and temperatures of the heat bath are $kT_1/(\hbar\Omega_1)=1$   and $kT_2/(\hbar\Omega_1)=0.1$, respectively.
}
\label{fig:l}
\end{figure}
\begin{figure}[h]
\includegraphics[scale=0.95]{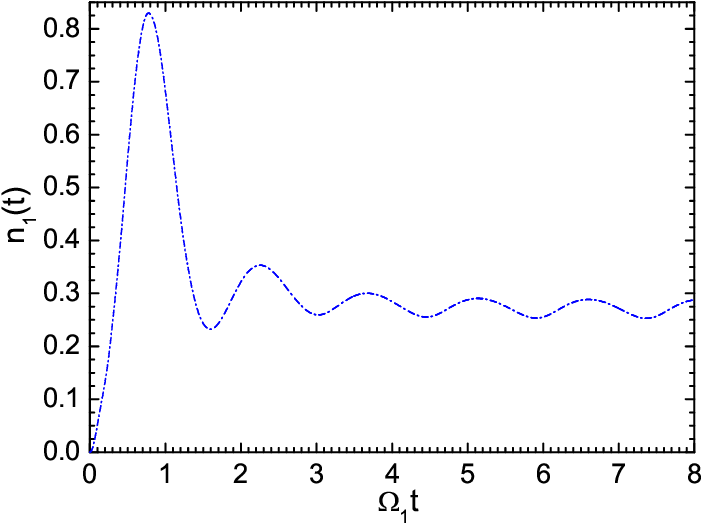}
\caption{
Calculated dependence of the average occupation number $n_1(t)$ on time $t$ for the bosonic oscillator.
The plot   corresponds to the  initially unoccupied, $n_1(0)=0$,  oscillator state.
In the calculations, the same parameters are used as in Fig. \ref{fig:l}.
}
\label{fig:2}
\end{figure}
\begin{figure}
\begin{center}
\includegraphics[scale=0.85,angle=0]{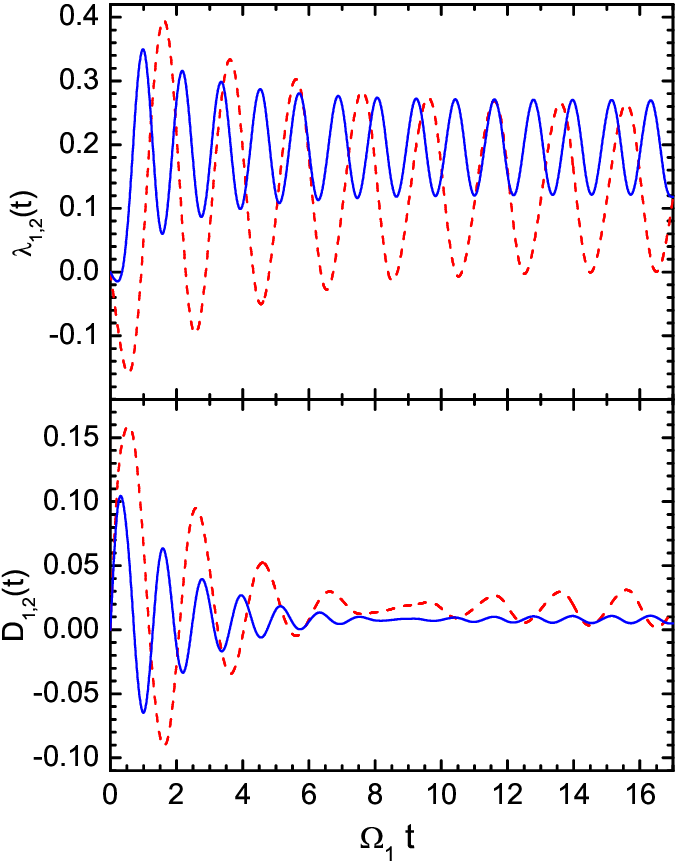}
\end{center}
\vspace{1cm}
\caption{
Calculated dependencies of the friction $\lambda_{1,2}(t)$    and  diffusion    $D_{1,2}(t)$   coefficients on time $t$ for the
two bosonic oscillators (dashed (the first oscillator)  and solid (the second oscillator)   lines).
The baths  have the same
level densities  with the Lorenzian  cut-off parameters $\gamma_1/\Omega_1=\gamma_2/\Omega_1=12$ and $\gamma_1/\Omega_2=\gamma_2/\Omega_2=6$.
Here the coupling strengths  between the oscillators and baths are the same, $\alpha_1=\alpha_2=0.03$,
temperatures $kT_{1}/(\hbar \Omega_{1})=kT_{2}/(\hbar \Omega_{1})=0.5$  and $kT_{1}/(\hbar \Omega_{2})=kT_{2}/(\hbar \Omega_{2})=0.25$, and
the ratio of renormalized frequencies of the bosonic oscillators is  $\Omega_{2}/\Omega_{1}=2$.
}
\label{fig:3}
\end{figure}
\begin{figure}[h]
\includegraphics[scale=0.8]{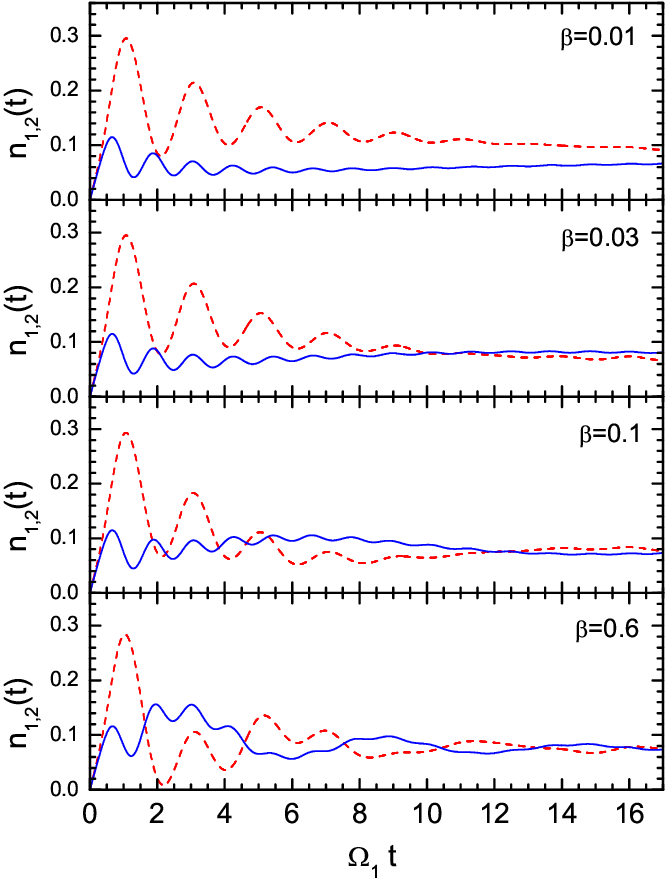}
\caption{
Calculated dependencies of the average occupation numbers $n_{1,2}(t)$ on time $t$
at different indicated coupling strengths $\beta$ between two bosonic oscillators
(dashed (the first oscillator)  and solid (the second oscillator)   lines).
The plots  correspond to the  initially unoccupied, $n_{1,2}(0)=0$,  oscillator states and $dn_{1,2}(0)/dt=0$.
In the calculations, the same parameters are used as in Fig. \ref{fig:3}.
}
\label{fig:4}
\end{figure}
As seen in Fig.~1, after quite a short transient time, $\Omega_1 t\le 0.5$
($\Omega_1$ is the renormalized frequency of oscillator (see Appendix A)),
the  $\lambda_1(t)$ and $D_1(t)$    oscillate  with the same period.
The ratio $D_1(t)/\lambda_1(t)$ has no asymptotic value and
is a periodic function at large time (Fig. 1).
This behavior at large time is due to the fact that the friction
 coefficient $\lambda_1(t)$ is a linear combination of friction
coefficients  in the system taken with either two bosonic thermostats ($\lambda_{\scriptstyle b}(t)$) or
two fermionic thermostats ($\lambda_{\scriptstyle f}(t)$). The friction (diffusion) coefficient of the system
with two bosonic thermostats oscillates as a function of time, while the friction
(diffusion) coefficient of the system with two fermionic thermostats
have an asymptotic limit (see Eqs. (\ref{lambda}), (\ref{Dif}), (\ref{lambdaA}), and (\ref{DifA}))  \cite{PRE2020n}.
However, the ratios $D_{{\scriptstyle  f}}^{(1)}/\lambda_{\scriptstyle  f}$ and $D_{{\scriptstyle  b}}^{(2)}/\lambda_{\scriptstyle  b}$
have asymptotic values $I^{(1)}_{\text{f}}$ and $I^{(2)}_{\text{b}}$, respectively (see Appendix A, Eq.~(A13)).
The ratio $D_1/\lambda_1$ is rewritten at large time ($t\to\infty$) as
$$\frac{D_1}{\lambda_1}\to\frac{\frac{I^{(1)}_{\text{f}}}{\lambda_{\scriptstyle  b}}+\frac{I^{(2)}_{\text{b}}}{\lambda_{\scriptstyle  f}}}{\frac{p-2I^{(1)}_{\text{f}}}{\lambda_{\scriptstyle  b}}+\frac{1-p}{\lambda_{\scriptstyle  f}}}.$$
Because $\lambda_{\scriptstyle  f}$ has an asymptotic, in general case this expression oscillates together with $\lambda_{\scriptstyle  b}$ at large time $t$.
In the case of system with two  heat baths of the same statistics, the ratio of diffusion and friction coefficients in the master-equation (\ref{eq:namaster2})
is constant at large times  (Appendix A) \cite{PRE2020n}. As demonstrated in Refs.~\cite{PRE2020n,PhysicaA2019,PRE2020}, in the case of heat baths of different statistics
the asymptotic ratio $D_1/\lambda_1$ becomes constant only at the specific condition
$$\frac{1}{1-p}I^{(2)}_{\text{b}}=\frac{\frac{1}{p}I^{(1)}_{\text{f}}}{1-\frac{2}{p}I^{(1)}_{\text{f}}}.$$
In this case we obtain Bose-Einstein distribution for $n_1(\infty)$  in the Markovian weak-coupling limit and
the same temperatures of both heat baths.
If the RWA coupling instead of FC is taken between the system and heat baths, there is also
asymptotic of $D_1/\lambda_1$ Ref.~\cite{JCP2013,PRE2020n,PhysicaA2019,PRE2020}.
Note that in the case of one heat bath there is always asymptotic occupation number and
our results are consistent with those obtained in Refs.~\cite{PRE1997,PRD1992,PRE2013,PRL2022,JCP2013,PRP1998,PRA2020}
based on the exact time evolution of the damped harmonic oscillator.

As a result of the lack of a stationary value of $D_1(t)/\lambda_1(t)$,
the  occupation number $n_1(t)$ oscillates around a certain average value at large times, so it has also no asymptotic limit (Fig. 2).
In Fig. 2, the  dependence of $n_1(t)$ on time $t$ is shown up to  time $t=8/\Omega_1$, but we numerically checked for non-stationarity of  $n_1(t)$ up to time $t>400/\Omega_1$
which is long enough to exclude possible transitional effects.
Note that the master-equation  (\ref{eq:namaster2})   was solved numerically with high accuracy.
At $\Omega_1t\ge 3$, the period of oscillations of $n_{1}(t)$   is found to be close to $2\pi/\Omega_1$.
This period is determined only by the properties of the system but not by the initial condition and the characteristics of heat baths  \cite{PRE2020n}.
The stable mode of  oscillations is determined by the energy balance, that is,
the equality of the energy lost by the oscillator to the heat baths and the energy supplied from the heat baths to the oscillator.
Therefore, the  occupation number oscillates independently of the heat baths, and the period of oscillations  carries
information about the system  \cite{PRE2020n}.
So we have an opportunity to control these oscillations by changing the oscillator frequency
and, correspondingly, to create a non-stationary memory storage device \cite{PRE2020n}.

The characteristics of our bosonic oscillator are very similar to those of the self-oscillating systems  \cite{Oni}.
The stable mode of self-oscillation is also determined by the energy balance, that is,
the equality between the energy lost by the system and the energy supplied from the external source to the oscillating system. However, our bosonic oscillator
takes energy from the heat baths and gives energy back to them. Our system is more resistant to the noise
because it directly takes into account the effects of environment.
Thus, one can say that the bosonic oscillator considered is a new kind of the self-oscillating system.
The results obtained in this section can be generalized as follows:
if the fluctuation-dissipative relation is not satisfied in a non-equilibrium system ($D_1(t)/\lambda_1(t)\ne$constant at large times),
then the occupation number  does not converge to a stationary value at $t\to\infty$.
From this point of view, a new self-oscillating  system is a non-equilibrium system,
in which the fluctuation-dissipative relation relation is not valid.

Note that in general the self-oscillating systems are nonlinear and of great interest from the point of view of both fundamental natural science and numerous applications.
There is no single approach to nonlinear methods,
but a number of effective methods have been developed for a wide range of applications \cite{Oni}.
However, their study faces significant difficulties in comparison with  systems  to which our system belongs.

\section{Two coupled bosonic oscillators linearly fully coupled to their own fermionic and bosonic  heat baths}
Let us consider two linearly coupled bosonic oscillators that are embedded in their own fermionic and
bosonic (fermionic-bosonic) heat baths.
The Hamiltonian of the total system
(two coupled quantum systems plus  four heat baths)  is written as
\begin{equation}
\label{ham2}
H=\sum_{k=1,2}\left[H_c^{(k)} + \sum_{\lambda=1,2} H_{\lambda}^{(k)} +  \sum_{\lambda=1,2}H_{c,\lambda}^{(k)}\right] + H_{12},
\end{equation}
where
\begin{equation}
\label{eq:coll2}
H_c^{(k)} = \hbar\omega_{(k)}a^{\dagger}_k a_k,
\end{equation}
\begin{equation}
\label{eq:baths2}
H_{ \lambda}^{(k)} = \sum_i \hbar \omega_{\lambda,i} c^{(k)\dagger}_{\lambda,i} c^{(k)}_{\lambda,i},
\end{equation}
\begin{equation}
\label{eq:intc12}
H_{c, \lambda}^{(k)} = \sum_i \alpha_{\lambda,i}^{(k)} (a^{\dagger}_k+a_k)(c^{(k)\dagger}_{\lambda,i}+c^{(k)}_{\lambda,i}),
\end{equation}
and
\begin{equation}
\label{eq:int12}
H_{12} = i\beta_0 (a^\dagger_1 a_2 - a_1 a^\dagger_2).
\end{equation}
Here, the terms $H_c^{(k)}$, $H_{ \lambda}^{(k)}$, $H_{c, \lambda}^{(k)}$ are the Hamiltonians  of the oscillator "$k$" [$a^\dagger_k$ ($a_k$)  are the creation (annihilation) operators],
heat bath "$\lambda$" [$c^{(k)\dagger}_{\lambda,i}$ ($c^{(k)}_{\lambda,i}$)  are the creation (annihilation) operators],
and  interaction between the oscillator "$k$" and heat bath "$\lambda$" with coupling constant $\alpha_{\lambda,i}^{(k)}$, respectively.
The term  $H_{12}$ is the coordinate-momentum interaction Hamiltonian between two oscillators \cite{M1,M8,Zub,Klim,Petruccione}.
The constant $\beta_0$ is the coupling constant between two bosonic oscillators.

Deriving the  equations of motion
for the operators
$c^{(k)\dagger}_{\lambda,i}(t)a_k(t)$,
$a_k(t)c^{(k)}_{\lambda,i}(t)$,  $a^{\dag}_{1}a_2$  and substituting their formal solutions in the equation of motion for the operators $a^\dagger_1(t) a_1(t)$,  $a^\dagger_2(t) a_2(t)$, $a^{\dag}_{1}(t)a_2(t)$,
averaging all operators over the whole system, and assuming the  independence of the heat bath oscillators and the mean-field approximation,
we obtain that the evolutions of occupation numbers $n_{1,2}(t)=\langle a^{\dag}_{1,2}a_{1,2}\rangle$
are determined by  the solution of the system of two coupled master-equations,
\begin{eqnarray}
\left\{
\begin{array}{l}
\displaystyle
\frac{d^2n_{1}(t)}{dt^2}+2\lambda_1(t)\frac{dn_{1}(t)}{dt}+2\frac{d\lambda_1(t)}{dt}n_{1}(t)+\beta[n_{1}(t)-n_{2}(t)]=2\frac{dD_1(t)}{dt}\\
\\
\displaystyle
\frac{d^2n_{2}(t)}{dt^2}+2\lambda_2(t)\frac{dn_{2}(t)}{dt}+2\frac{d\lambda_2(t)}{dt}n_{2}(t)+\beta[n_{2}(t)-n_{1}(t)]=2\frac{dD_2(t)}{dt}
\end{array}
\right. ,
\label{eq:namaster2-2}
\end{eqnarray}
where the constant  $\beta$  is related to the coupling strength between two bosonic oscillators
with the renormalized frequencies $\Omega_{1,2}$ (see Appendix A), and the values of $\lambda_{1,2}(t)$ and $D_{1,2}(t)$
are the time-dependent friction and diffusion coefficients for the first and second oscillators, respectively.
The value of $\beta$ is taken independent of time and provides the energy exchange between them.
At $\beta=0$, we obtain two decoupled master-equations (\ref{eq:namaster3}).
In Fig. 3, the values of  $\lambda_{1,2}(t)$ and $D_{1,2}(t)$ oscillate out of the phase.
When interacting, the two bosonic oscillators influence each other since the connection between them is carried out in both directions.

The system of master-equations for $n_{1,2}(t)$  (\ref{eq:namaster2-2})  is solved numerically with high accuracy.
Since the friction and diffusion coefficients in Eq.~(\ref{eq:namaster2-2}) are functions of time,
the dynamics of $n_{1,2}(t)$  is not linear.
As seen in Fig. 4,  after   short transient time,  the  occupation numbers $n_{1,2}(t)$ oscillate and have no asymptotic limits.
In the cases of
$\beta=$0.1 and 0.6, the values of $n_{1,2}(t)$ oscillate in the anti-phase  at large time, $\Omega_1t\approx$7--17, that is, the energy exchange occurs between the systems.
At $\beta=$0.01 and 0.03,  the anti-phase oscillations occur  at time greater than $17/\Omega_{1,2}$.
The oscillation amplitudes and frequencies change with time $t$
due to the coupling between the oscillators. These changes are more pronounced at larger coupling strength  between the oscillators (Fig. 4).
The influence of the characteristics of thermal reservoirs of different statistics on the
periods of oscillations of $n_{1,2}(t)$ [$2\pi/\Omega_{1,2}$] is almost negligible. This makes the self-oscillating systems quite robust and
one can control the asymptotic oscillations of both bosonic oscillators by changing their frequencies $\Omega_{1,2}$ and coupling constant $\beta$.





The main feature of synchronization is that the ratio of the characteristic time scales of interacting systems, which were independent in the absence of coupling,
 is rational number    \cite{Oni}.    Because the    periods   of oscillations of
occupation numbers $n_{1,2}(t)$ at large times are almost the same, the system of two coupled bosonic oscillators is similar to two coupled self-oscillating systems
which have the property of synchronization. This effect  is also generalized to   a   finite number of coupled bosonic oscillators.
Synchronization for a   finite number of coupled bosonic oscillators
can be used to increase the stability of quantum computers with a dynamic or non-stationary  memory  \cite{PRE2020n}.

\begin{figure}
\begin{center}
\includegraphics[scale=0.85]{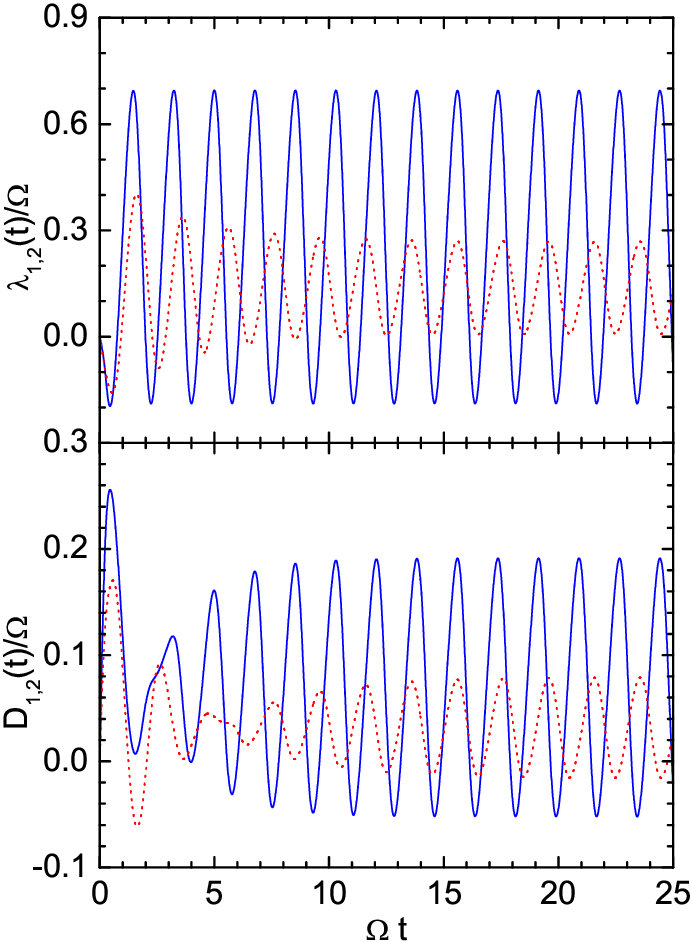}
\end{center}
\vspace{1cm}
\caption{
Calculated dependencies of the friction $\lambda_{1,2}(t)$    and  diffusion    $D_{1,2}(t)$   coefficients on time $t$ for
two bosonic oscillators (dotted (the first oscillator coupled with fermionic-bosonic baths)  and solid
(the second oscillator coupled to bosonic-bosonic baths)   lines). The renormalized frequencies of the bosonic oscillators are $\Omega_{1}=\Omega_{2}=\Omega$.
The fermionic-bosonic baths  have the same
level densities  with the Lorenzian cut-off parameters $\gamma_{1f}/\Omega=\gamma_{1b}/\Omega=12$
and temperatures $kT_{1f}/(\hbar \Omega)=0.1$ and $kT_{1b}/(\hbar \Omega)=1$, respectively.
The bosonic-bosonic baths  have the
level densities  with the cut-off parameters
  $\gamma_{2b}/\Omega=12$ and 15, respectively, and  temperatures are $kT_{2b}/(\hbar \Omega)=1$ and 0.1, respectively.
The coupling strengths  between the first oscillator  and fermionic-bosonic baths  are the same, $\alpha_{1f}=\alpha_{1b}=0.03$.
The coupling strengths  between the second oscillator  and bosonic-bosonic baths  are  $\alpha_{2b}=0.05$ and $0.03$, respectively.
}
\label{fig:5}
\end{figure}
\begin{figure}[h]
\includegraphics[scale=1]{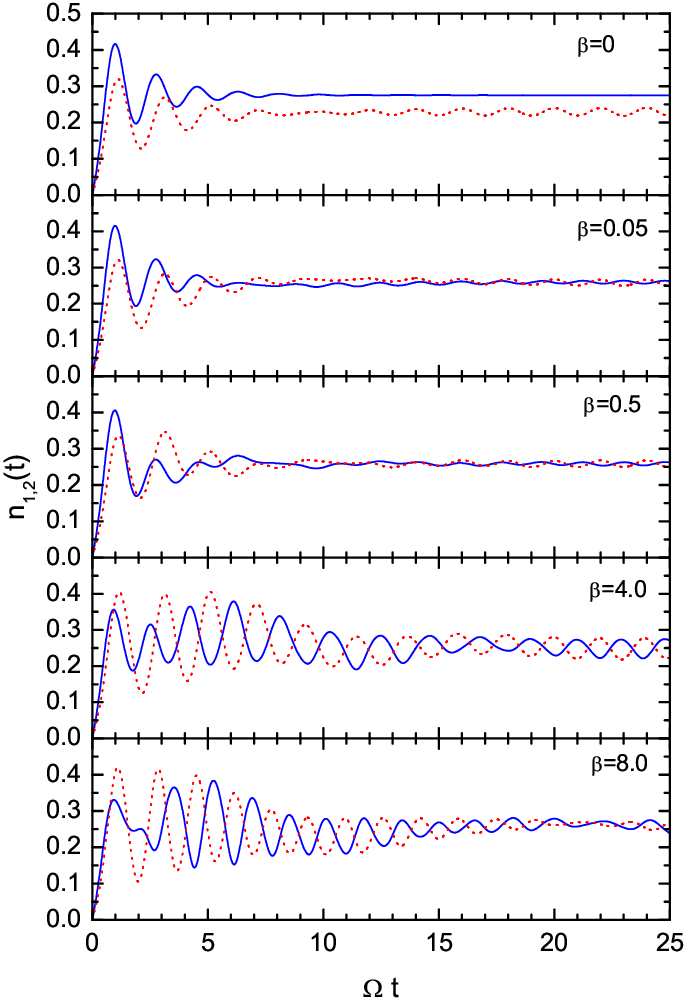}
\caption{
Calculated dependencies of the average occupation numbers $n_{1,2}(t)$
on time $t$ at  indicated coupling strengths $\beta$ between two coupled bosonic oscillators (dotted  (the first oscillator plus fermionic-bosonic baths)  and
solid (the second oscillator plus bosonic-bosonic baths)   lines).
The plots  correspond to the  initially unoccupied, $n_{1,2}(0)=0$,  oscillator states and $dn_{1,2}(0)/dt=0$.
In the calculations, the same parameters are used as in Fig. \ref{fig:5}.
}
\label{fig:6}
\end{figure}
\begin{figure}[h]
\includegraphics[scale=1]{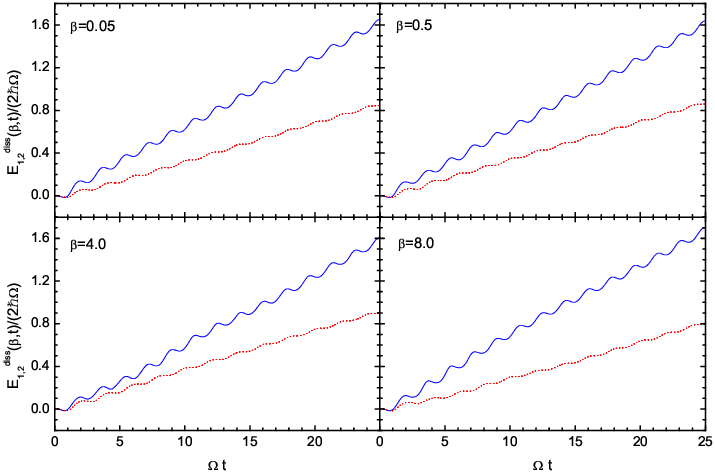}
\caption{
Calculated dependencies of the dissipation energies $E_{1,2}^{\rm diss}(\beta,t)$
on time $t$ at different indicated coupling strengths $\beta$ between two bosonic oscillators,
one of which (2) is coupled to bosonic-bosonic baths (solid lines) and the other (1) is connected with
fermionic-bosonic baths (dotted lines).
In the calculations, the same parameters are used as in Fig. \ref{fig:5}.
}
\label{fig:7}
\end{figure}
\begin{figure}[h]
\includegraphics[scale=1.2]{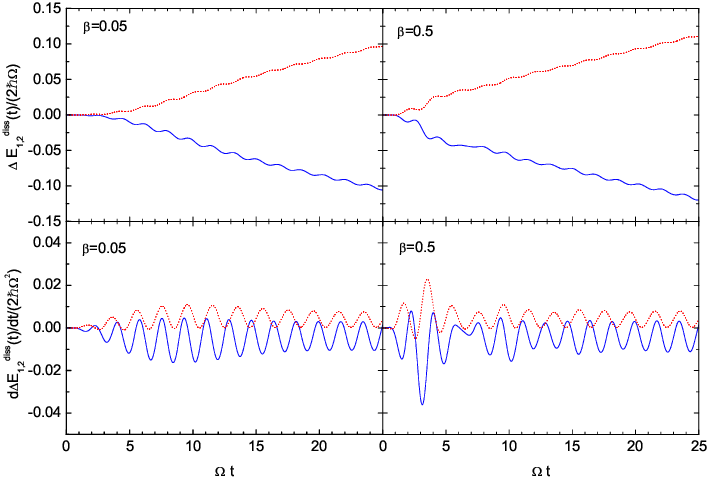}
\caption{
Calculated dependencies of the values of $\Delta E_{1,2}^{\rm diss}(t)$ and $d\Delta E_{1,2}^{\rm diss}(t)/dt$
on time $t$ at different indicated coupling strengths $\beta$ between two bosonic oscillators,
one of which is coupled to bosonic-bosonic baths (2, solid lines) and the other is connected with
fermionic-bosonic baths (1, dotted lines).
In the calculations, the same parameters are used as in Fig. \ref{fig:5}.
}
\label{fig:8}
\end{figure}

\section{Interaction between two bosonic oscillators  connected respectively with
fermionic-bosonic  and  bosonic-bosonic baths}

Let us consider the system with two coupled bosonic oscillators of
the same renormalized frequencies $\Omega_1=\Omega_2=\Omega$.
The first (second) oscillator is linearly fully coupled to
fermionic-bosonic (bosonic-bosonic)  baths. So, when these oscillators are decoupled,
the occupation number $n_{1}(t)$ of the first  oscillator has no asymptotics  at large times,
while  the occupation number $n_{2}(t)$ of the second oscillator   has asymptotics ($n_2(\infty)=D_2(\infty)/\lambda_2(\infty)$).
The evolutions of occupation numbers $n_{1,2}(t)$ are determined by
the solution of the system of two coupled master-equations (\ref{eq:namaster2-2}).
In Fig.~\ref{fig:5},  the values of friction $\lambda_{1,2}(t)$ and diffusion $D_{1,2}(t)$
coefficients are shown for two systems.
 The values of diffusion and friction  coefficients equal zero at $t=0$.
After quite a short transient time,
the  $\lambda_{1,2}(t)$ and $D_{1,2}(t)$    oscillate  as a function of $t$ with the same frequency.
The ratio $D_1(t)/\lambda_1(t)$
is a periodic function of time at large times and has no asymptotic value.
The ratio $D_2(t)/\lambda_2(t)$ is constant at large times.

As seen in Fig.~\ref{fig:6}, after quite a short transient time, $\Omega t\le 1$,
the  occupation numbers $n_{1,2}(t)$ start to oscillate  with the same frequencies ($\Omega t \approx 2$) and decreasing amplitude.
The amplitude of the oscillations is greater for a large coupling constant $\beta$ between the oscillators.
At large times, $n_{1,2}(t)$ oscillate in the anti-phase  around a certain average value,
although at the beginning they are almost in phase.
The phase shift occurs faster with   increasing  coupling constant $\beta$.

As a result of the absence of a stationary value of $D_1(t)/\lambda_1(t)$ and, correspondingly, $n_{1}(t)$,
the first and second oscillators   become  the sources of energy.
As seen in Fig.~\ref{fig:7}, the dissipation energies
\begin{eqnarray}
E_{1,2}^{\rm diss}(\beta,t)=2\hbar\Omega\int_0^tdt'\lambda_{1,2}(t')n_{1,2}(t')
\end{eqnarray}
of the first and second oscillators   increase  oscillating with time. The oscillators lose and gain energy due to friction and diffusion, respectively.
The values of $E_{1,2}^{\rm diss}(\beta,t)$ weakly depends on $\beta$ and  $E_{2}^{\rm diss}(\beta,t)>E_{1}^{\rm diss}(\beta,t)$.
The   $E_{2}^{\rm diss}(\beta,t)$ increases over the time interval $\Omega t\approx 1$
than slightly decreases over the time interval $\Omega t\approx 0.5$ (Fig.~\ref{fig:7}).
It looks like an experimental observation of oscillating chemiluminescene emission during Belousov-Zhabotinsky reaction
(reduction of bromate by malonic acid) when tris(2,2'-bipyridine) ruthenium (II), Ru(bpy)$^{2^+}_3$, was used as a catalyst \cite{Balz}.
As seen in Fig. 8, the dissipation energy at $\beta\neq 0$  relative to the energy at
$\beta=0$ ($\Delta E_{1,2}^{\rm diss}(t)=E_{1,2}^{\rm diss}(\beta\neq 0,t)-E_{1,2}^{\rm diss}(\beta=0,t)$)
behaves differently in the case of the first and second oscillators.
In the case of the first oscillator, this energy $\Delta E_{1,2}^{\rm diss}(t)$ decreases, but in the case of the second oscillator, it increases.
On the other hand, the occupation numbers and, accordingly, the energies of the systems oscillate, which means that energy is exchanged between the oscillators.
The dissipated energy of the first oscillator increases due to $\beta\neq 0$ that can be considered as an energy source.
So the baths and coupling between the oscillators provide the dissipated energy.
Moreover, the absolute average dissipation energy rate  $d\Delta E_{1,2}^{\rm diss}(t)/dt$ is greater for the second oscillator than for the first one (Fig. 8).
Thus, the coupling of the two oscillators reduces the energy loss in the second oscillator.

\section{Summary}
It is shown  that a bosonic oscillator embedded in both fermionic and bosonic heat baths is a
new type of self-oscillating system where the energy is supplied from the heat baths to the oscillating system and back, and
the fluctuation-dissipative relation relation is not satisfied.
The fermionic  oscillator linearly fully coupled to
the  fermionic and  bosonic heat baths would have the same properties \cite{PRE2020n}.
The oscillator embedded in both fermionic and bosonic heat baths can be used as an energy source,
if it is possible to take away the energy of asymptotic oscillations fueled by heat baths.

In the case of two linearly coupled bosonic oscillators embedded in their own fermionic and bosonic heat baths,
the absence of equilibrium asymptotic  of occupation numbers was predicted.
At large times, the periods of oscillations of occupation numbers mainly depend
on the  frequencies of the corresponding oscillators  and, accordingly, carry
information about both oscillators. This is analogous to the synchronization effect of the self-oscillating system.
Similar behavior of the system of two coupled bosonic (fermionic) oscillators can be generalized to a system with an
arbitrary finite number of coupled bosonic (fermionic) oscillators.

\section*{Acknowledgments}

V.V.S. acknowledges the support of the Alexander von Humboldt-Stiftung (Bonn).
This work was partly supported by the DFG (Bonn, Grant No. Le439/16).


%




\newpage

\appendix

\section{Bosonic (Fermionic) oscillator coupled with two bosonic (fermionic) heat baths}
 Let us consider the case when two heat baths and  oscillator (with frequency $\omega$)  are either all  bosonic   or  all fermionic.
For these systems, details of
the derivation of the master-equation for the occupation number $n_{\text{a}}(t)$ of oscillator
are given in Ref. \cite{PRA2017}.
Here, using the notations $\text{a}=\text{f}$  and $\text{a}=\text{b}$ for fermionic and bosonic
oscillators, we directly write the final expression for
the master-equation ($\varepsilon_\text{f}=-1$ and $\varepsilon_\text{b}=1$)
\begin{equation}
\label{dnt2}
\frac{dn_{\text{a}}(t)}{dt}=-2\lambda_{\text{a}}(t)n_{\text{a}}(t)+2D_{\text{a}}(t),
\end{equation}
where
\begin{equation}
\label{lambdaA}
\lambda_{\text{a}}(t)=-\frac{1}{2}\frac{d}{dt}\ln\left[|A(t)|+\varepsilon_{\text{a}}|B(t)|^2\right]
\end{equation}
and
\begin{eqnarray}
\label{DifA}
D_{\text{a}}(t)&=&\sum_{\lambda=1,2}D^{(\lambda)}_{\text{a}}(t)=\lambda_{\text{a}}(t)\big[|B(t)|^2+I_{\text{a}}(t)\big]+\frac{1}{2}\frac{d}{dt}\big[|B(t)|^2+I_{\text{a}}(t)\big], \nonumber\\
D^{(\lambda)}_{\text{a}}(t)&=&\lambda_{\text{a}}(t)\big[J^{(\lambda)}(t)+I^{(\lambda)}_{\text{a}}(t)\big]+\frac{1}{2}\frac{d}{dt}\big[J^{(\lambda)}(t)+I^{(\lambda)}_{\text{a}}(t)\big]
\end{eqnarray}
are the time-dependent friction and diffusion coefficients, respectively.  Here, $\lambda_{\text{a}}(t=0)=D_{\text{a}}(t=0)=0$.
For the time-dependent coefficients $A(t)$, $B(t)$, $ M(w,t)$, $N(w,t)$,
the following expressions are obtained:
\begin{widetext}
\begin{equation}
\label{eq:dynamicp}
\begin{split}
A(t)&= i\sum_{k=1}^{4} \xi_ke^{s_kt}(s_k-s_0)(s_k-i\omega)(s_k+i\omega)^{-1}
[\alpha_1\gamma_1^2(s_k+\gamma_2)+\alpha_2\gamma_2^2(s_k+\gamma_1)],\\
 B(t)&= i\sum_{k=1}^{4} \xi_ke^{s_kt}(s_0-s_k)[\alpha_1\gamma_1^2(s_k+\gamma_2)+\alpha_2\gamma_2^2(s_k+\gamma_1)],\\
N(w,t)&=  \sum_{k=0}^4\xi_ke^{s_kt} (is_k-\omega)(s_k+\gamma_1)(s_k+\gamma_2),\\
M(w,t)&= -\sum_{k=0}^4\xi_ke^{s_kt} (is_k+\omega)(s_k+\gamma_1)(s_k+\gamma_2),
\end{split}
\end{equation}
\end{widetext}
where
\begin{equation}
\xi_k=\prod_{i=0,\\i\neq k}^4\frac{1}{s_k-s_i},
\end{equation}
 $s_0=-iw$, and $s_1$, $s_2$, $s_3$, $s_4$
 are the roots of the following equation:
\begin{equation}
\label{eq:polB}
(s^2+ \omega \Omega)(s+\gamma_1)(s+\gamma_2)+2s\omega[\alpha_1 \gamma_1(s+\gamma_2)+\alpha_2 \gamma_2(s+\gamma_1)]=0.
\end{equation}
In Eq. (\ref{eq:polB}), $\Omega=\omega-2 \alpha_1\gamma_1-2 \alpha_2\gamma_2$ is the renormalized frequency of oscillator.

In Eqs.~(\ref{lambdaA}) and (\ref{DifA}) for the time-dependent friction  and diffusion coefficients we use
the expression \cite{PRA2017}
\begin{eqnarray}
J^{(\lambda)}(t)&=&|B_{\lambda}(t)|^2+\frac{1}{2}\big[B_{1}(t)B^*_{2}(t)+B^*_{1}(t)B_{2}(t)\big],\nonumber\\
\label{eq:dynamic00001}
\end{eqnarray}
which results from the decomposition of  $B(t)$ as
\begin{eqnarray}
B(t)&=&\sum_{\lambda=1,2}B_{\lambda}(t), \nonumber\\
B_{1,2}(t)&=&i\alpha_{1,2} \gamma^2_{1,2}\sum_{k=1}^{4} \xi_ke^{s_kt} (s_0-s_k)(s_k+\gamma_{2,1}).
\label{eq:dynamic00002}
\end{eqnarray}

The following decomposition
\begin{eqnarray}
\label{eq:dyna12}
I_{\text{a}}(t)=\sum_{\lambda=1,2}I^{(\lambda)}_{\text{a}}(t),
\end{eqnarray}
\begin{eqnarray}
\label{eq:i1}
I^{(\lambda)}_{\text{a}}(t)=\frac{\alpha_\lambda\gamma_\lambda^2}{\pi}\int_0^{\infty}dw \frac{w}{\gamma_\lambda^2+w^2}\left[n^{eq}_\lambda(w,T_\lambda)|M(w,t)|^2+[1+\varepsilon_\lambda n^{eq}_\lambda(w,T_\lambda)] |N(w,t)|^2\right] \nonumber\\
\end{eqnarray}
is used in Eq. (\ref{DifA}).
In Eq. (\ref{eq:i1}),
$n^{eq}_{\lambda}(w,T_\lambda)=(\exp[\hbar w/(kT_{\lambda})]-\varepsilon_\lambda)^{-1}$  are
equilibrium Fermi-Dirac (Bose-Einstein)  distributions of the fermionic (bosonic) heat baths "$\lambda$".
The $T_{\lambda}$ is the initial thermodynamic temperature  of the corresponding heat   bath.
Here, we  introduce the spectral density $\rho_\lambda (w)$ of the
heat-bath excitations, which allows us to replace the sum over  $i$ by the integral over the frequency $w$:
$\sum_i...\to \int_{0}^{\infty}dw\rho_\lambda (w)...$.
For all baths, we consider the following spectral function with the Lorenzian cutoffs \cite{M1,Petruccione,Armen}:
\begin{eqnarray}
\dfrac{\alpha_{\lambda,i}^2}{\hbar ^2 w_{\lambda,i}}\to\dfrac{\rho_\lambda (w) \alpha_{\lambda,w}^2}{\hbar^2 w}=\dfrac{1}{\pi }\alpha_{\lambda}\dfrac{\gamma_{\lambda}^2}{\gamma_{\lambda}^2+w^2},
\label{eq_222zxc}
\end{eqnarray}
where the memory time $\gamma_{\lambda}^{-1}$ of  dissipation is inverse to the bandwidth of the heat-bath
excitations which are coupled to the collective  system.
The relaxation time of the heat-bath should be much less than   the
characteristic collective time, i.e.,  $\gamma_{\lambda}\gg \omega$.

In the case of two heat baths of the same statistics,
from Eq. (\ref{dnt2}) we obtain
the asymptotic occupation numbers
\begin{eqnarray}
\label{eq:asym1a1}
n_{\text{a}}(t\rightarrow \infty)=\frac{D_{\text{a}}(t\rightarrow \infty)}{\lambda_{\text{a}}(t\rightarrow \infty)}=\sum_{\lambda=1,2}I^{(\lambda)}_{\text{a}}(t\rightarrow \infty)
\end{eqnarray}
 are defined only by the integrals \cite{PRA2017}
\begin{eqnarray}
I^{(\lambda)}_{\text{a}}(t\rightarrow \infty)=\frac{\alpha_\lambda\gamma_\lambda^2}{\pi}\int_0^{\infty}dw\frac{w(\gamma_{\bar\lambda}^2+w^2)\left\{[\omega+w]^2 n^{eq}_{\lambda}(w,T_{\lambda})+[\omega-w]^2[1+\varepsilon_\lambda n^{eq}_{\lambda}(w,T_{\lambda})]\right\}}{(s_1^2+w^2)(s_2^2+w^2)(s_3^2+w^2)(s_4^2+w^2)}. \nonumber\\
\end{eqnarray}
If $\lambda=1$, then ${\bar\lambda}=2$ and vise-versa.
In the Markovian  limit (weak-couplings constants $\alpha_\lambda$  and high temperatures $T_\lambda$) and at $T_1\neq T_2$,   we obtain
\begin{eqnarray}
n_{\text{a}}(t\rightarrow \infty)=p n^{eq}_{1}(\omega,T_1)+(1-p)n^{eq}_{2}(\omega,T_2),
\end{eqnarray}
where
$$p=\frac{\alpha_1}{\alpha_1+\alpha_2}.$$
If $T_1=T_2=T$, then
\begin{eqnarray}
n_{\text{a}}(t\rightarrow \infty)=n^{eq}_{\text{a}}(\omega,T)
\end{eqnarray}
 has the usual form of Fermi-Dirac or Bose-Einstein distribution (a thermal equilibrium).

\end{document}